# Conditions for the Trivers-Willard hypothesis to be valid: A minimal population-genetic model

N. V. Joshi

*Center for Ecological Sciences, Indian Institute of Science, Bangalore 560 012, India*

E-mail - nvjoshi@ces.iisc.ernet.in

## Abstract

The very insightful Trivers-Willard hypothesis, proposed in the early 1970s, states that females in good physiological conditions are more likely to produce male offspring, when the variance of reproductive success amongst males is high. The hypothesis has inspired a number of studies over the last three decades aimed at its experimental verification, and many of them have found adequate supportive evidence in its favour. Theoretical investigations, on the other hand, have been few, perhaps because formulating a population-genetic model for describing the Trivers-Willard hypothesis turns out to be surprisingly complex.

The present study is aimed at using a minimal population genetic model to explore one specific scenario, viz. how is the preference for a male offspring by females in good condition altered when 'g', the proportion of such females in the population changes from a low to a high value. As expected, when the proportion of such females in good condition is low in the population, i.e., for low values of 'g', the Trivers-Willard (TW) strategy goes to fixation against the equal investment strategy. This holds true up to gmax, a critical value of 'g', above which the two strategies coexist, but the proportion of the TW strategy steadily decreases as 'g' increases to unity. Similarly, when the effect of well-endowed males attaining disproportionately high number of matings is more pronounced, the TW strategy is more likely to go to fixation. Interestingly, the success of the TW strategy has a complex dependence on the variance in the physiological condition of females. If the difference in the two types of conditions is not large, TW strategy is favoured, and its success is more likely as the difference increases. However, beyond a critical value of the difference, the TW strategy is found to be less and less likely to succeed as the difference becomes larger. Possible reasons for these effects are discussed.

___________________________________________________________________________

## Introduction

Ever since the appearance of the classic paper by Hamilton in the late sixties (Hamilton, 1967), sex ratio theory has been an active research area in the field of theoretical evolutionary biology. The two other notable advances were mainly due to Robert Trivers. The Trivers-Hare (Trivers and Hare, 1976) article introduced the now famous 1:3 investment ratio between male: female offspring in social insects and led to a flurry of theoretical and experimental research papers arguing for and against the role of haplodiploidy in social evolution, on the basis of the theoretically predicted and experimentally observed sex investment ratios in social insects. The other article, the Trivers-Willard hypothesis (Trivers and Willard, 1973), suggested that in societies where the variance in reproductive success of males is large, females in 'good condition' should preferentially invest in male offspring. A number of studies have demonstrated the validity of this hypothesis in many different systems, ranging from house mice (Meikle and Thornton, 1995), white-tailed deer (Burke and Birch, 1995) and elks (Smith, Robbins and Anderson, 1996) to even spinach (Freeman et al, 1994). The impact of this publication went beyond the natural sciences, and spawned a number of studies in social sciences as well, covering for example human populations in modern Venezuela (Chacon-Puignau and Jaffe, 1996), the Hungarian Gypsies (Bereczkei and Dunbar, 1997), reproductive success of elite males in nineteenth century Germany (Klindworth and Voland, 1995) and even the offspring sex ratios observed in the families of the presidents of the United States of America (Betzig and Weber, 1995).

While the aspects of sex ratio investment discussed in the first two papers have generated a large number of theoretical investigations, there have been only a few attempts at modeling the Trivers-Willard hypothesis (Charnov et al 1981, Charnov 1982, Anderson and Crawford 1993, Leimar 1996, see Godfray and Werren 1996), despite the simplicity of its underlying logic. Trivers and Willard pointed out that there is generally a large difference in the ***variance*** of the reproductive success between males and females. By and large, most females would be able to produce a few offspring. On the other hand, a small number of males often father a disproportionately large number of offspring, while many males have little or no access to females and hence have very poor reproductive success. The males enjoying high reproductive success are most often the most dominant ones, as judged by size, body condition and other related attributes such as large horns, tusks, feathers etc. For those species for whom the above criteria are satisfied, if a female is in good physiological condition, and if her condition is likely to carry over to her son, then it would make sense to produce a son and invest in him, thereby increasing the probability of he being one of the dominant males. On the other hand, a female in not-so-good a condition would be better off by producing daughters; these are likely to have at least a moderate reproductive success, but a not-so-well-endowed son will probably have none. Simply stated, the Trivers-Willard hypothesis predicts that females in good condition will preferentially invest in sons and those not in such a good condition would bias the investment in favour of daughters.

Though the qualitative predictions are almost self-evident, a more detailed examination of the hypothesis reveals that it is not so easily amenable to conventional population genetic modeling. Firstly, one of the key features of the hypothesis is that the 'good condition' of the mother is passed on to the son. Clearly, this is not at all the conventional kind of inheritance, and cannot be modeled as a heritable trait. Secondly, the relative proportions of the females in different body conditions are likely to influence the sex investment decisions to a considerable extent. If only a small number of females are in good condition, the hypothesis in its present form does seem to make a reasonable prediction. On the other hand, if a majority of them are in a good condition, the variance in the reproductive success of the males will either decrease, or the chance of a specific male enjoying a high reproductive success will be low. Biasing investment in favour of males does not necessarily seem to be an advantageous strategy under such circumstances. What, then, is the range of validity of the Trivers-Willard hypothesis? A quantitative formulation of the hypothesis is essential for answering such a question under the scenario described above. This is the objective of the present investigation.

## The Model

An attempt is made to set up as simple a population genetic model as possible which will cover all the essential features of the Trivers-Willard hypothesis. A haploid, panmixing, infinite population is assumed, and the sex investment bias is assumed to be governed by a single locus. The different features and parameters of the model are described below.

### *Heterogeneity in the body condition of females in the population*

It is assumed that females can exist in only two types of body conditions: good and normal. The proportion of females in good condition is denoted by 'g'; the proportion of females in normal condition is then seen to be equal to (1-g). Further, a unit resource is assumed to be available for females in normal condition to invest in offspring. The females in good condition are on the other hand assumed to have 'R' units of resources for the same purpose, and $R > 1$.

### *The strategies*

To begin with, only two alternative strategies are considered, each coded for by an allele. The 'F' allele (Fisherian, coding for equal investment in the two sexes, as per R.A. Fisher's classic explanation as to why the sex ratio is generally 1:1 in most circumstances) makes its bearer produce an equal number of male and female offspring. The 'T' allele (Triversian) bearing females produce only sons when they are in good condition and an equal proportion of sons and daughters when they are in a normal condition.

In fact, these strategies can be implemented in several different ways. For example, the resource R may be invested equally in male and female offspring by the F strategy, with each sex obtaining R/2 units of resource. Alternatively, the whole resource R may be invested either fully in sons or fully in daughter, with probability 0.50 for the two situations. This later approach is what is implemented in the present model, and forms one of the critical assumptions of the model (see discussion).

### *Enhancement in male reproductive success due to maternal investment*

As explained above, Triversian females in good body condition are expected to invest the whole of 'R' into males. Under the assumptions of the Trivers-Willard hypothesis, this is supposed to considerably enhance the competitive ability of the males, and hence the probability of high reproductive success. This situation is modeled by assuming that 'R' units in males are equivalent to $R^d$ males being produced. The parameter 'd' (indicative of disparity between competitive ability of males) can vary between 1 and infinity. The situation d=1 corresponds to the case where there is no *additional enhancement*, whereas at the other extreme, d=infinity suggests that only those males born to females in good condition are able to sire offspring in the next generation.

### *Dynamics of allelic frequencies*

To begin with, the population is assumed to consist entirely of the F allele, corresponding to the Fisherian, equal investment strategy. This is invaded by the T allele, corresponding to the 'Triversian', body-condition-dependent strategy, initially present in extremely low frequency. For given values of the three parameters of the model viz. 'g', the proportion of females in good body condition; 'R', the excess resource available to such females and 'd', the parameter translating this extra investment into enhanced competitive ability of males, the changes in the allelic frequencies of 'T' and 'F' are computed. Details of the various steps involved in computing the changes in allelic frequencies are described in the Appendix; a brief outline is presented below.

Changes in the frequencies of inseminated females (of the two genotypes and phenotypes) are computed from one generation to the next. The two variables of interest are 'x', the proportion of females of genotype 'T' and 'y', the proportion of sperms of the same 'T' genotype, as seen from the proportions of females inseminated by the 'T' genotype. The calculations are carried out only up to first order, and terms involving squares of x and y, and the product term xy are omitted. As can be seen from the appendix, this makes it possible to express (x', y'), the frequencies of the two variables of interest in the next generation, as a linear combination of (x, y), the values of these variable in the present generation. The coefficients of x and y are function of the three parameters of the model, viz. g, R and d. If the dominant eigenvalue of the 2X2 transition matrix relating the frequencies (x', y') to (x, y) is greater than unity, the proportion of the 'T' allele increases from one generation to the next, and it is thus able to 'invade' the resident population corresponding to the 'F' allele.

The same exercise is repeated by making 'T' as the resident allele and 'F' as the invading allele. If the dominant eigenvalue of the transition matrix is less than unity in this case, it implies that the 'F' allele is unable to invade the 'T' population. Under these circumstances, where the T allele is able to invade F, and is stable against invasion by F, one can conclude that the Trivers-Willard hypothesis has been validated, since the T allele outcompetes the normal, equal-investment, 'F' allele. In the present investigation, an attempt has been made to identify the regions of validity of the Trivers-Willard hypothesis in the three-dimensional space of the parameters g, R and d.

The results of the stability analysis based on computation of eigenvalues were confirmed by carrying out detailed simulations taking into account all the terms (i.e., not just the ones up to first order). All the calculations were carried out in double precision using Fortran programs.

## Results

The most important parameter of the model is 'g', the proportion of females in a good physiological condition. For very small values of 'g', and for almost any values of 'R', the resource available to such females, and 'd', the disparity parameter, the Trivers-Willard (T) strategy is seen to go to fixation in the population. In fact, the population of (F), Fisherian (equal investment) strategy is ***always*** invadable by T, regardless of the value of g. At low values of g, T is univadable by F. However, as g increases and reaches a critical value (see Fig. 1), T becomes invadable by F, and the two strategies coexist stably. As g is further increased, the proportion of T genotypes in the population gradually comes down (Fig. 1), eventually reaching zero as g reaches unity. The four curves in the figure refer to four different values of R, viz. 1.1, 1.2, 1.3 and 1.4, and it is seen that a higher value of R in this range is favourable to the T strategy, since it leads to a higher level of equilibrium frequency of T. A similar dependence of equilibrium frequency of T on g is seen in Figure 2, where the curves are for different values of 'd', the disparity parameter. It is seen that as d increases from 1.5, to 3 to 5 to 9, the situation becomes more favourable to T.

One of the important aspect of the two figures is the presence of gmax, a critical value of 'g', up to which the T strategy goes to fixation. As seen from figure 1, this value is seen to increase with R, as R increases from 1.1 to 1.4. Similarly, Fig. 2 shows that gmax increases as the value of d, the disparity parameter increases.

A more detailed dependence of gmax on R, the high quantity of resource available to females in good condition is shown in Fig. 3. For low values of R, gmax indeed increases with R, indicating that a higher level of resource favours the T strategy. However, as R is increased still further, the value of gmax is seen to begin to decrease, reaching very low levels for high values of R. In other words, beyond a certain point, a high level of resource for females in good condition actually seem to be unfavourable to the Trivers-Willard hypothesis. This apparently counterintuitive result, however, seems to have a plausible explanation, described in the discussion section. The different curves in Fig.3, corresponding to different values of d, the disparity parameter, show that high values of d favour the T strategy, as expected.

This is seen more clearly in Fig. 4, which depicts the variation of gmax as a function of d. As expected, gmax increases with increasing values of d. However, the rate of increase slows down, and gmax seems to reach saturation for high values of d. The four curves in the figure correspond to four different values of R, viz., R=1.5, 3, 5 and 9, and also bring out the point made in the previous paragraph that high values of R in fact become less favourable to T.

## Discussion

The minimal population genetic model considered above demarcates the range of parameter values under which the Trivers-Willard hypothesis would be valid. The quantitative results are in accordance with what one expected qualitatively using verbal arguments. Thus a low proportion of females in good condition, a higher level of R, the resource enjoyed by the females in good condition and a high value of the disparity parameter, which converts this extra resource into competitive ability all favour the Trivers-Willard strategy. The only apparently counterintuitive result is that as R increases, conditions become less favourable to the T strategy. The explanation of this apparent anomaly lies in the magnitude of contribution to the next generation from the F strategy, and the situation is similar to one where the value of g is high. In other words, even if g is low, if R is high enough to make gXR attain a high value, then there would be a relatively large number of well endowed F males, and consequently, the advantage enjoyed by the T males is not as high. Secondly, the F strategy attains a higher fitness via the daughters born of females in good condition, thus further enhancing the relative fitness of F compared to the T strategy, which produces no daughters when in good a condition. As a consequence of these two effects, the T strategy becomes less favourable as R increases to very high values.

The saturation of gmax with increasing d is also an expected result. As d increases, the well-endowed males become more and more effective in converting the extra resource into competitive ability, to such an extent that almost all the matings are due to well-endowed males. Once this situation has been attained, there are no more matings to be obtained, and thus there is no further advantage available to the T strategy, resulting in the saturation of gmax at high values of d.

The model described so far in this investigation examined only one part of the Trivers-Willard hypothesis, viz. that well-endowed females should preferentially produce sons. The other part of the hypothesis states that poorly endowed females should preferentially produce daughters. To examine this aspect, the counterpart of the minimalist model described above, namely, a population consisting only of a mixture of females in normal condition and in poor condition was also studied. The T strategy in this case would correspond to the females in poor condition producing only daughter. The results of the model are by and large similar to the one described above, and predictions based on both the aspects of the Trivers-Willard hypothesis are thus seen to be borne out by the minimal population genetic model.

A more general model, which includes all the three physiological conditions (good, normal and poor) has also been formulated and a few exploratory investigations have been carried out. As expected, the full model shows that a much higher region in the parameter space is favourable to the T strategy.

As mentioned earlier (while describing the strategies in the section 'The model' above), a critical assumption of the model is that for the F strategy, females when in good physiological condition, i.e., with resource R, produce either a son or a daughter with equal probability. An equally plausible assumption would have been that they produce both sons and daughters with an investment R/2 in each of them. In such a scenario, however, the sons born to F mothers even in good condition will not have as high a competitive ability as those born of T mothers, since the later will have an investment of R, and due to the disparity parameter d which enhances this difference, the T strategy will turn out to be even more advantageous than what has been seen in the present model. The focus of the current investigation, however, was to examine the conditions for the success of the T strategy even under conditions not readily favourable to it; hence the assumption of full investment of R in the offspring and equal probability of sons and daughters was implemented for the F strategy in the current investigation.

One of the limitations of the present analysis is that an explicit Evolutionarily Stable Strategy (ESS) has not been derived. It is possible that unlike the simple only-sons-or-only-daughters strategy considered above, a more complex strategy with differential quantitative investment in sons and daughters under the two body conditions would turn out to be an ESS. Studies along these lines, as well as exploration of more complex scenarios (the general model described above, evolution when more than two strategies present at a time, effect of diploidy, dominance and linkage, stochastic effects in small populations) are in progress.

## Acknowledgements

Financial assistance from Ministry of Environment and Forests, Government of India is gratefully acknowledged. I thank J.A. Santosh, Dhrubajyoti Naug, H.S. Arathi and Anindya Sinha for the intense discussions which triggered the formulation of this population-genetic model, and Prof. H. Sharat Chandra, whose constant encouragement was responsible for its development.

This paper is dedicated to the memory of W.D. Hamilton, whose untimely demise has robbed evolutionary biology of its rightful place in Stockholm.

# Appendix

The population is assumed to be haploid, sexual and panmictic, and consisting predominantly of the 'F' allele, being 'invaded' by a small proportion of 'T' allele. The changes in frequencies from one generation to the next are computed for the proportions of inseminated females of the two genotypes.

Let x be the proportion of females of genotype 'T', and 'y' be the proportion of sperms of genotype T (where x and y are nearly equal to zero). The corresponding proportions of the 'F' genotype are given by (1-x) and (1-y) respectively. The proportions of the various genotypes of inseminated females are then given by (where the first letter denotes the genotype of the female and the second one, of the sperm)

FF = (1-x) (1-x) = 1-x-y   to first order

FT = (1-x) y     = y   to first order

TF = x (1-x)     = x   to first order and

TT = xy          = 0   to first order.

A proportion 'g' in each category are assumed to be in good condition, and (1-g) in a normal body condition. As explained in the text, the investment in offspring available to the females of the two conditions is R and 1, respectively. The number and type of offspring produced by the eight categories of inseminated females of various genotypes and phenotypes are given below.

FF females produce both sons and daughters of genotype F. Those in good condition produce either a son or a daughter with equal probability, each endowed with a resource R units, while those in normal condition produce either a son or a daughter with an equal probability, each endowed with a unit resource. The contribution of the FF females to the next generation is then equivalent to $(g.R^d+(1-g))/2$ males of genotype F and $(gR+(1-g))/2$ females of genotype F.

FT females produce both sons and daughters, of both the T and F genotypes. Those in good and in normal conditions behave as described above, and the contribution to the next generation is $(g.R^d+(1-g))/4$ males each of genotypes F and T, and $(gR+(1-g))/4$ females each of genotypes F and T.

TF females also produce both sons and daughters of both T and F genotypes. Those in good condition, however, produce *only sons*, with an investment of R units. Those in normal condition produce both sons and daughters with equal probability, each with an investment of one unit. The contribution to the next generation then is $(gR^d+(1-g))/4$ males each of genotypes T and F and $(1-g)/4$ females each of genotypes T and F.

For a linear stability analysis of the situation where T invades a pure F population, only the above six categories need to be considered. However, to carry out the simulation of changes in genotypic frequencies in the full nonlinear mode, the two categories (TT females in good and normal conditions) are needed. These are also needed for the stability analysis of the situation where a pure T population is invaded by F.

TT females produce both sons and daughters of genotype T. Those in good condition produce only sons, endowed with a resource R units, while those in normal condition produce either a son or a daughter with an equal probability, each endowed with a unit resource. The contribution of the TT females to the next generation is then equivalent to $(g.R^d + (1-g)/2)$ males of genotype T and $(1-g)/2$ females of genotype T.

Using the above expressions and details, the relative frequencies of the females in the next generation can be easily worked out. Under the first-order approximation, the frequencies x' and y', of females and males respectively of T genotype in the next generation are expressed as functions of expressions involving R, d and g, and the frequencies x and y in the present generation. In fact, the expression for a vector (x', y') can be written down as a product of a transition matrix and the vector (x, y). The elements of the transition matrix are functions of the three parameters of the model g, R and d. The eigenvalues of the matrix can be computed for any set of values of g, R and d. If the dominant eigenvalue is greater than unity, T is able to successfully invade the population.

A very similar set of expressions and procedure is used for examining the invadability of a pure T population by the F allele.

The same expressions described above, but without making the first order approximation (i.e., including all the terms involving products xy, and all the eight categories of inseminated females) were used to simulate the changes in the genotypic frequencies from one generation to the next. The simulations were carried out primarily to ensure that the results obtained by the linear stability analysis remain valid under the full nonlinear treatment. Simulations were also essential to obtain the equilibrium frequencies of the two strategies when stability analysis predicted their coexistence.

# Legends to Figures

Figure 1.
Variation in the equilibrium frequency of 'T' as a function of g, the proportion of females in good physiological condition for different values of R, the reproductive resource of the females in good condition. The four curves correspond to R=1.1 (solid diamond), R=1.2 (solid square), R=1.3 (solid triangle) and R=1.4 (cross).

Figure 2.
Variation in the equilibrium frequency of 'T' as a function of g, the proportion of females in good physiological condition for different values of d, the disparity parameter indicating the efficiency of males in converting the higher resource into extra competitive ability. The four curves correspond to d=1.5 (solid diamond), d=3.0 (solid square), d=5.0 (solid triangle) and d=9.0 (cross).

Figure 3.
Variation in gmax, the maximum value of g at which the T strategy can go to fixation, as a function of R. The four different curves correspond to different values of d; d=1.5 (solid diamond), d=2.0 (solid square), d=3.0 (solid triangle) and d=5.0 (cross).

Figure 4.
Variation in gmax, the maximum value of g at which the T strategy can go to fixation, as a function of d. The four different curves correspond to different values of R; R=1.5 (solid diamond), R=3.0 (solid square), R=5.0 (solid triangle) and R=9.0 (cross).

**Figure 1**

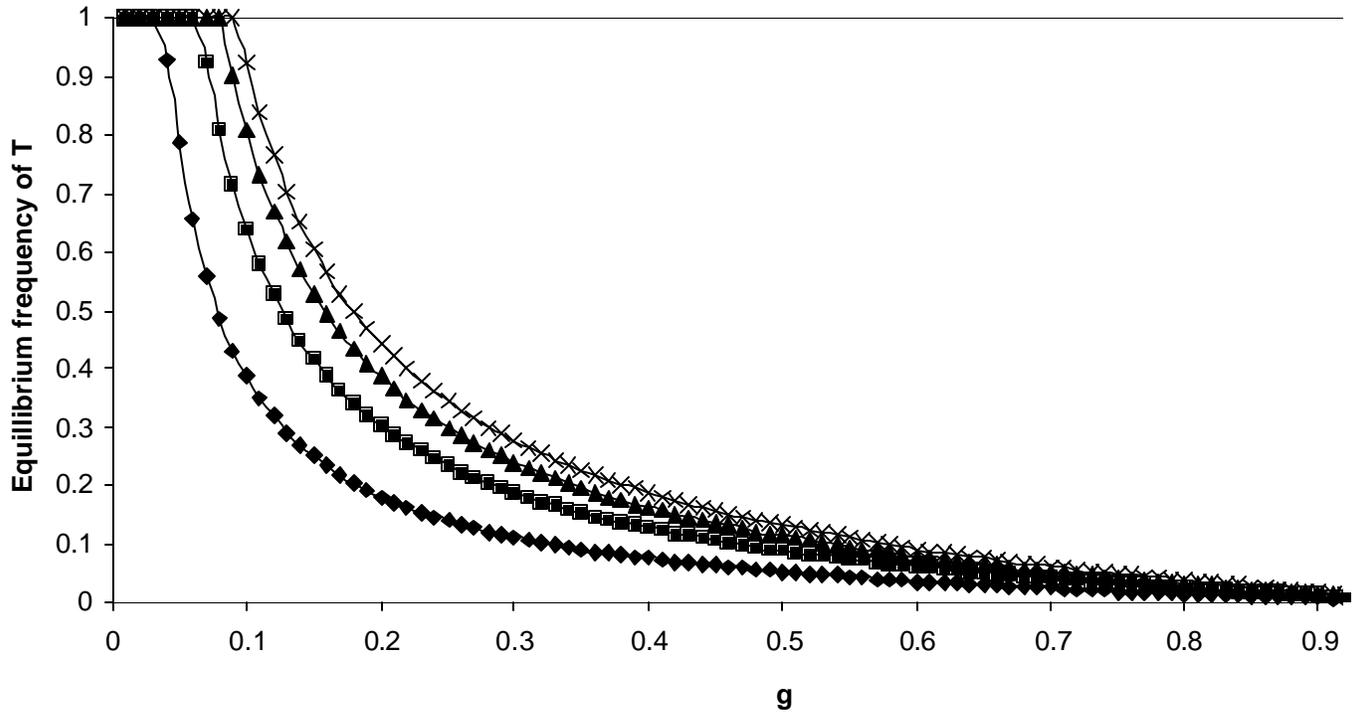

**Figure 2**

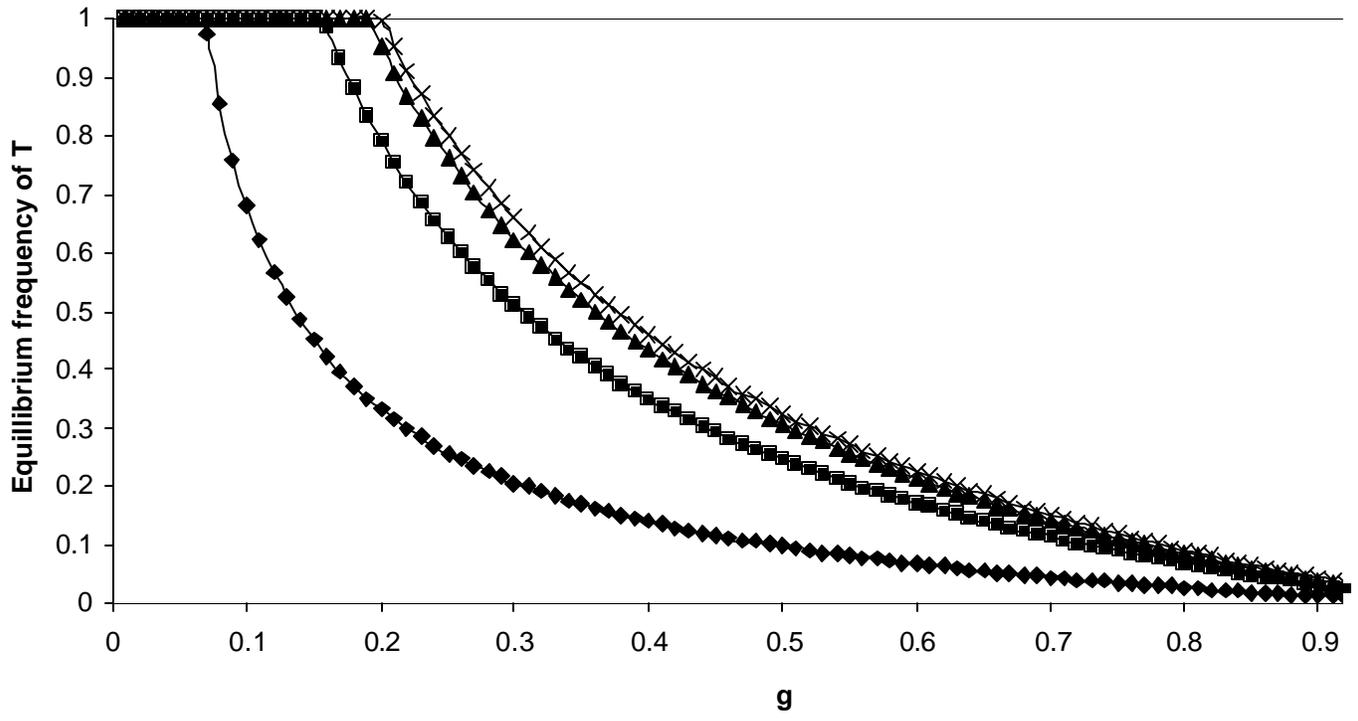

**Figure 3**

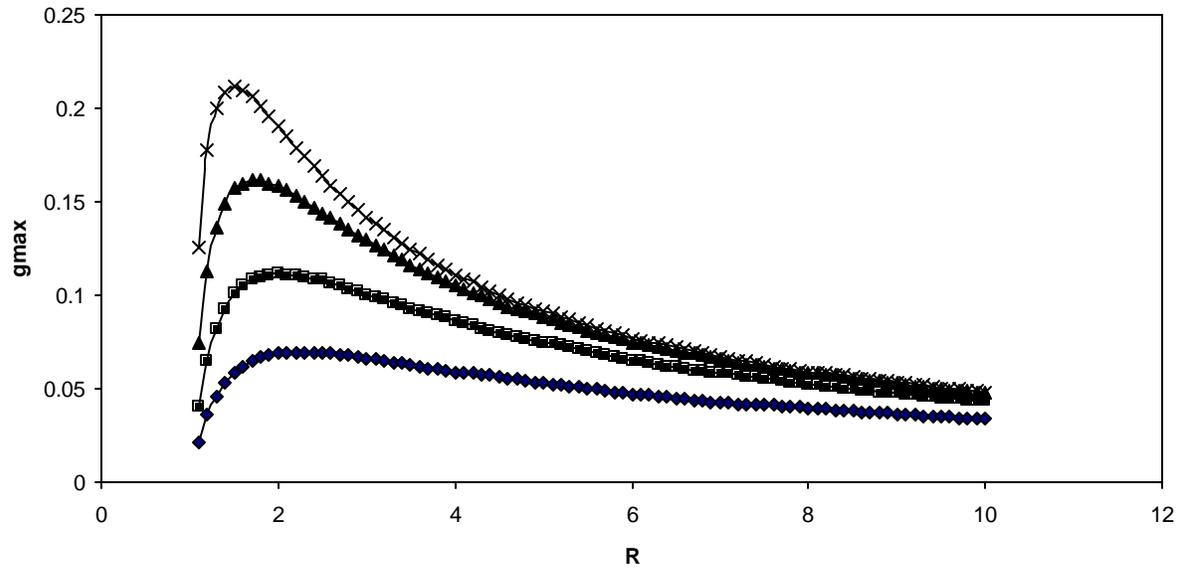

**Figure 4**

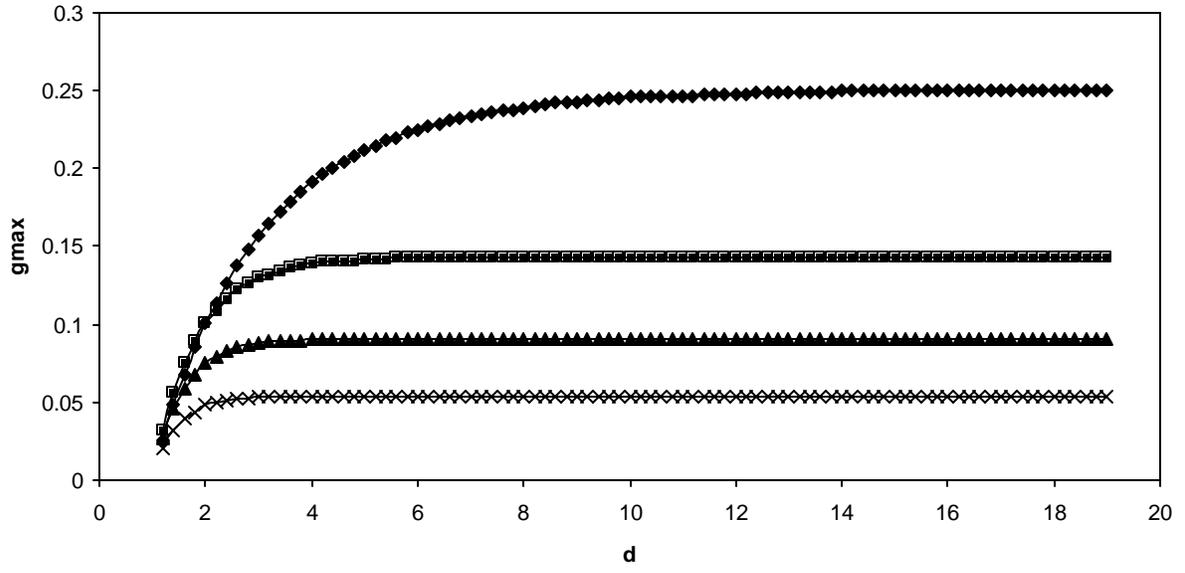